\documentstyle[12pt,epsf]{article}

\renewcommand{\thefootnote}{\fnsymbol{footnote}}
\newcommand{\newsection}{
\setcounter{equation}{0}
\section}

\def\appendix#1{
  \addtocounter{section}{1}
  \setcounter{equation}{0}
  \renewcommand{\thesection}{\Alph{section}}
  \section*{Appendix \thesection\protect\indent \parbox[t]{11.715cm} {#1}}
  \addcontentsline{toc}{section}{Appendix \thesection\ \ \ #1}
  }

\newcommand {\defeq}{\stackrel{\rm def}{=}}
\newcommand{\tr}[1]{\:{\rm tr}\,#1}

\def\e{{\,\rm e}\,}

\def\eop{\vspace*{\fill}\pagebreak}
\newcommand{\rf}[1]{(\ref{#1})}
\newcommand{\eq}[1]{Eq.~(\ref{#1})}
\def\be{\begin{equation}}
\def\ee{\end{equation}}
\def\beq{\begin{equation}}
\def\eeq{\end{equation}}
\def\bea{\begin{eqnarray}}
\def\eea{\end{eqnarray}}

\def\LA{\left\langle}
\def\RA{\right\rangle}

\def\A{{\cal A}}
\def\U{{\cal U}}

\def\cl{{\rm cl}}
\font\mybbs=msbm10 at 9pt
\def\bbs#1{\hbox{\mybbs#1}}
\font\mybb=msbm10 at 12pt
\def\bb#1{\hbox{\mybb#1}}

\def\IT{{\bb T}}
\def\ITs{{\bbs T}}

\def\IR{{\bb R}}
\def\IZ{{\bb Z}}

\def\IZs{{\bbs Z}}
\def\I1{{\bb I}}
\def\I1s{{\bbs I}}

\newcommand{\ra}{\rightarrow}
\begin{document}

\begin{titlepage}
\begin{flushright}
ITEP--TH--47/00\\
hep-th/0009028\\
July, 2000
\end{flushright}
\vspace{1cm}

\begin{center}
{\LARGE Reduced Models and \\[.4cm]
 Noncommutative Gauge Theories}\\
\vspace{1.4cm}
{\large Yuri Makeenko}\footnote{E--mail:
makeenko@itep.ru \ \ \ \
 makeenko@nbi.dk \ } \\
\vskip 0.2 cm
{\it Institute of Theoretical and Experimental Physics,}
\\ {\it B. Cheremushkinskaya 25, 117259 Moscow, Russia}
\\ \vskip .1 cm
and  \\  \vskip .1 cm
{\it The Niels Bohr Institute,} \\
{\it Blegdamsvej 17, 2100 Copenhagen {\O}, Denmark}
\end{center}
\vskip 1.5 cm
\begin{abstract}

This is a short review of the relation between the
Reduced Models and Noncommutative Yang--Mills Theories (NCYM)
based on the works~\cite{ANMS99,ANMS00a,ANMS00b}.

\vspace{6pt}
Contents:
\begin{itemize}\vspace{-4pt}
\addtolength{\itemsep}{-4pt}
\item[1.] Twisted Eguchi--Kawai model (TEK),
\item[2.] Mapping onto NCYM,
\item[3.] Morita equivalence,
\item[4.] Fundamental matter,
\item[5.] Wilson loops in NCYM,
\item[6.] D-brane interpretation.
\end{itemize}
\vspace{3pt}

Talk given at the 11th International Seminar ``Quarks'2000'',  
Pushkin, Russia, May 13--21, 2000 
and the E.S.~Fradkin Memorial Conference, Moscow, June 5--10, 2000.

\end{abstract}

\end{titlepage}
\setcounter{page}{2}
\renewcommand{\thefootnote}{\arabic{footnote}}
\setcounter{footnote}{0}

\newsection{Introduction}

The recent interest in noncommutative gauge theories has been inspired by
the work~\cite{CDS98} on compactification of Matrix 
Theory~\cite{BFSS96,IKKT97}. Matrix Theory is generically of the type of
reduced models~\cite{EK82} where (infinite) matrices are space-independent 
while the space-dependence appears when expanding around a classical vacuum. 
In order for a reduced model to be equivalent to the 't~Hooft limit of
a large-$N$ quantum field theory on the continuum space-time, it should 
be either quenched~\cite{BHN82} or twisted~\cite{GO83a,EN83,GO83b,GAKA83}.  

It has been recently recognized~\cite{AIIKKT} that there exists 
a limit of the twisted reduced models when they describe quantum field
theories on the noncommutative space represented by operators $\hat x^\mu $
with the commutation relation
\beq
\left[ \hat x^\mu, \hat x^\nu \right]= i\theta^{\mu\nu} \,.
\label{commutation}
\eeq
The multiplication of the fields is given by the star-product
\beq
\phi _1 (x) \ast \phi _2 (x) \defeq
\phi _1 (x)\,\e^{\frac i{2} \,
\overleftarrow{\partial_\mu} \theta_{\mu\nu}
\overrightarrow{\partial_\nu}}\,
\phi _2 (x) 
\label{starproduct}
\eeq
while the action of noncommutative $U_\theta(1)$ theory is
\beq
S= \frac{1}{4e^2}\int {\cal F}^2
\eeq
with ${\cal F}$ being the
noncommutative field strength
\beq
{\cal F}_{\mu\nu}=\partial_{\mu} \A_\nu -\partial_{\nu} \A_\mu -
i \left(\A_\mu \ast\A_\nu- \A_\nu \ast\A_\mu  \right).
\label{defF}
\eeq 
The description~\cite{GO83b} of the planar limit of the $U(N)$ Yang--Mills
theory is associated
with $\theta\ra\infty$. This relation between the twisted reduced models 
and noncommutative gauge theories has been further elaborated in
Refs.~\cite{IIKK,BM,ANMS99,ANMS00a,ANMS00b}. 
In particular, the proper observables
of noncommutative gauge theories have been constructed.

I review in this talk the relation between the twisted
reduced models and noncommutative gauge theories. 
A particular emphasis is given to equivalences between some
noncommutative and ordinary (or ``commutative'') gauge theories
known as Morita equivalence~\cite{AS}. The simplest example 
is the above-mentioned equivalence of the large-$\theta$ limit of
 noncommutative $U_\theta(1)$ gauge theory and the large-$N$ limit of
ordinary $U(N)$ Yang--Mills. Another example is that
noncommutative $U_\theta(1)$ gauge theory in a box with periodic boundary
conditions is  equivalent at some rational values of $\theta$ to
ordinary Yang--Mills theory in a smaller box with twisted boundary conditions
representing the 't~Hooft flux. These results are obtained
starting from the twisted reduced model at finite $N$ which is
mapped onto noncommutative gauge theory on a lattice of finite 
spatial extent thereby providing rigorous results for a
regularized quantum theory. An extension of Morita equivalence to
the case when fundamental matter is incorporated is described using
the proper twisted reduced model~\cite{Das83}. The path-integration over
the fundamental matter
determines observables in noncommutative gauge theory which are expressed
via both closed and open Wilson loops. A description of Morita equivalence
as T-duality in the D-brane language in presented.

\newsection{Twisted Eguchi--Kawai model}

\subsection{The definition \protect{\cite{GO83b}}}

The twisted Eguchi--Kawai model (TEK) is built out of $D$ $N\times N$
unitary matrices $U^{ij}_\mu$ ($\mu=1,\ldots,D$). The partition function
\beq
Z_{\rm TEK}= \int \prod_\mu dU_\mu \e^{\frac1{2g^2} \sum_{\mu\neq\nu} 
Z^*_{\mu\nu}
\tr{U_\mu U_\nu U^\dagger_\mu U^\dagger_\nu} + {\rm h.c.}}
\label{TEK}
\eeq
is of the type of Wilson's lattice gauge theory on a unit hypercube
with twisted boundary condition.
The twisting factor $Z_{\mu\nu}$ is
\beq
Z_{\mu\nu}=\e^{4\pi i n_{\mu\nu }/N} \in \IZ_N~~~~~~
({\rm integral}~~n_{\mu\nu}=- n_{\nu\mu})
\label{Zmunu}
\eeq
where we assume $N$ to be odd.

The twisted Eguchi--Kawai model possesses the symmetries
\bea
{\rm gauge}&:&~~U_\mu \ra  \Omega U_\mu \Omega^\dagger \,, 
\label{gauge} \\
\IZ_N^D &:&~~U_\mu \ra Z_\mu U_\mu ~~~\left(Z_\mu \in \IZ_N\right).
\label{ZD}
\eea

The vacuum state is given modulo a gauge transformation by
\beq
U_\mu^\cl = \Gamma_\mu
\label{vacuum}
\eeq
where $\Gamma_\mu$ are twist eaters obeying the Weyl--'t~Hooft commutation 
relation
\beq
\Gamma_\mu\Gamma_\nu = Z_{\mu\nu} \Gamma_\nu\Gamma_\mu\,.
\label{WT}
\eeq
They are known explicitly for any $n_{\mu\nu}$. 

The simplest twist is when
\beq
n_{\mu\nu}= L^{D/2-1} \varepsilon_{\mu\nu}\,,~~~~\varepsilon_{\mu\nu}=
\left(
\begin{array}{rrrrrr}
0 & +1& & & &\\
-1& 0 & & & &\\
 & &0 & +1& &\\
 & &-1& 0 & &\\
 && &.& & \\
 && && &. \\
\end{array}
\right)
\label{twist}
\eeq
and $N=L^{D/2}$. The group $SU(N)$ can then be decomposed in the direct
product $SU(N)\supset\prod_1^{D/2} \otimes SU(L)$ 
so that $\Gamma_i$, $\Gamma_{i+1}$
($i=1,\ldots,D/2$) can be chosen to be Weyl's clock and shift matrices
for one of $SU(L)$'s. Then $\Gamma_\mu^L=1$ for this simplest twist.

\subsection{Continuum limit of TEK \protect{\cite{GAKA83}}}

The continuum limit of TEK is reached when the lattice spacing 
$a\ra0$ ($N\ra\infty$) so that
\beq
U_\mu=\e^{iaA_\mu}\,,~~~~\Gamma_\mu =\e^{ia\gamma_\mu}\,,
\label{defU}
\eeq
where $A_\mu$ and $\gamma_\mu$ are Hermitean.
The relation~\rf{WT} turns into the Heisenberg commutator
\beq
\left[\gamma_\mu, \gamma_\nu \right] = i B_{\mu\nu} \,,~~~~~~~~
B_{\mu\nu} = \frac{4\pi n_{\mu\nu}}{Na^2}\,.
\label{H}
\eeq

The action of continuum TEK becomes
\beq
S=\frac 1{4g^2} \tr { \left( \left[A_\mu, A_\nu \right] -iB_{\mu\nu} \right)^2}
\eeq
and the vacuum configuration reads
\beq
A^\cl_\mu = \gamma_\mu
\eeq
modulo a gauge transformation $A_\mu\ra  \Omega A_\mu\Omega^\dagger$. 
The Wilson loops of large-$N$ Yang--Mills theory are represented by
\beq
W(C)=\left\langle \frac 1N \tr {\rm P}\e^{-i \int_C d\xi^\mu \gamma_\mu}
\frac 1N \tr {\rm P}\e^{i \int_C d\xi^\mu A_\mu} 
\right\rangle_{\rm TEK}\,.
\label{WL}
\eeq
They are nontrivial since $A_\mu$'s do not commute. The first trace on
the right-hand side of \eq{WL} vanishes for open loops. For a closed loop
it represents the flux of $B_{\mu\nu}$ through a surface bounded 
by the contour.

\subsection{Compactification of reduced models}

A compactification to a D-torus $\IT^D$ can be described~\cite{CDS98}
by imposing on $A_\mu$ the quotient condition
\beq
A_\mu+2\pi R_\mu \delta_{\mu\nu}  = \Omega_\nu A_\mu \Omega_\nu^\dagger 
\label{quotient}
\eeq
where $\Omega_\nu$ are unitary transition matrices.
Taking the trace of \eq{quotient}, we see
that a solution exists only for infinite matrices (= Hermitean operators).

Exponentiating $A_\mu$ according to \eq{defU} with a dimensionful parameter
$a$, we get
\beq
\e^{2\pi i a \delta_{\mu\nu} R_\mu} U_\mu = 
\Omega_\nu U_\mu \Omega_\nu^\dagger
\label{quotientU}
\eeq
where $U_\mu$ is unitary. This equation is a $N\times N$ 
matrix discretization
of \eq{quotient} and has solutions (described below) for finite $N$.

Taking the trace of \eq{quotientU}, we conclude
that $U_\mu$ should be traceless which is the case for the twist eaters. 
Taking the determinant of \eq{quotientU}, we conclude that 
$a R_\mu N $ should be integral. The consistency of \eq{quotientU}
also requires 
\beq
\Omega_\mu \Omega_\nu = z \,\Omega_\nu \Omega_\mu 
\eeq
with $z \in \IZ_N$. The quotient condition~\rf{quotientU} is compatible
with the gauge symmetry~\rf{gauge} if $\Omega$ commutes with
the transition matrices $\Omega_\nu$'s.

\subsection{Finite-N solution~\protect{\cite{ANMS99}}}

To describe a solution to \eq{quotientU}, let us first introduce 
the Weyl basis on $gl(N)$:
\beq
J_k= \Gamma_1^{k_1} \cdots \Gamma_D^{k_D} 
\e^{2\pi i \frac 1N \sum_{\mu>\nu} n_{\mu\nu}k_\mu k_\nu} ,
\label{defJ}
\eeq
where the last factor provides a symmetric product and $J_{L-k}=J^\dagger_k$.
These generators obey 
\be
J_k J_q = J_{k+q} \e^{2\pi i \frac 1N \sum_{\mu,\nu} k_\mu n_{\mu\nu} q_\nu} 
\label{JJ}
\ee 
which results finally in noncommutativity.

Let us choose
\be
\Omega _\mu = \prod_\nu \Gamma_\nu^{m \varepsilon_{\mu\nu}}
\ee
where $m$ is an integer. Then a particular solution to \eq{quotientU}
is given by
\be
U_\mu^{(0)} = \Gamma_\mu
\ee
while a general solution is 
\be
U_\mu =V_\mu \Gamma_\mu\,,
\ee
where $V_\mu $ obeys
\be
V_\mu =\Omega_\nu V_\mu \Omega_\nu^\dagger \,.
\label{tildequotient}
\ee
The solution to \eq{tildequotient} reads
\beq
V_\mu^{ij} = \sum_{k\in \IZs_m} \left(J_k^n\right)^{ij} U_\mu(k)
\label{tildeU}
\eeq
where $n=L/m$ is an integer. Here $k$ runs from $1$ to $m$
since $\Gamma_\mu^L=1$. This $V_\mu$ obviously commutes with
$\Omega_\nu$.

Given the c-number coefficients $U_\mu(k)$ which describe dynamical
degrees of freedom, we can make a Fourier
transformation to get the field
\be
\U_\mu(x)= \sum_{k\in \IZs_m} \e^{2\pi i \frac{kx}{am}} U_\mu(k)
\label{U(x)}
\ee
which is periodic on a $m^D$ lattice (or equivalently on a discrete
torus $\IT^D_m$). The spatial extent of the lattice is therefore $l=am$.
The field $\U_\mu(x)$ describes the same degrees of freedom as the 
(constraint) $N\times N$ matrix
$U^{ij}_\mu$ while the unitarity condition $U_\mu U^\dagger_\mu=1$
can be rewritten as
\beq
\U_\mu(x) \star \U^*_\mu(x) =1 \,,
\eeq
where $\U^*_\mu$ stands for complex conjugation 
and the lattice star-product is given by
\be
f(x)\star g (x) = \sum_{x,y} \e^{2 i (\theta^{-1})_{\mu\nu} y_\mu z_\nu}
f(x+y) g (x+z) 
\label{lstar}
\ee
with 
\be
\theta_{\mu\nu}= \frac{a^2mn}\pi \varepsilon_{\mu\nu}. 
\label{theta}
\ee
These formulas follow
from comparing expansions~\rf{tildeU} with \rf{U(x)} and using~\eq{JJ}.
As $a\ra0$, \eq{lstar} recovers \eq{starproduct} for the star-product in
the continuum.

\newsection{Mapping onto NCYM}

The twisted Eguchi--Kawai model~\rf{TEK} (in general with the quotient
condition~\rf{quotientU}) can be identically rewritten as a noncommutative
$U_\theta(1)$ lattice gauge theory. Given the relations~\rf{tildeU} 
and \rf{U(x)} 
between matrices and fields, we rewrite the action of TEK as
\be
S= \frac{1}{2e^2} \sum_{x\in \ITs^D_m} \sum_{\mu\neq\nu} 
\U_\mu(x)\star \U_\nu(x+a\hat\mu)
\star \U^\dagger_\mu(x+a\hat\nu)\star \U^\dagger_\nu(x)\,,
\label{NCLGT}
\ee
where $\hat \mu$ is a unit vector in the direction $\mu$
and the coupling constant $e^2=g^2 N$.
Analogously, the (constraint) measure $dU_\mu$ turns into the Haar measure
\beq
\prod _\mu dU_\mu \Rightarrow \prod _{x,\mu} d \,\U_\mu(x) \,.
\eeq
The action~\rf{NCLGT} is invariant under the star-gauge transformations
\be
\U_\mu(x) \ra \omega(x)\star \U_\mu(x)\star\omega^*(x+a\hat \mu) \,,
\label{stargauge}
\ee
where $\omega(x)$ is star-unitary 
($\omega\star\omega^*=\omega^*\star\omega=1$).
\eq{stargauge} is the counterpart of~\eq{gauge}.

The usual TEK (without the quotient condition) is associated
with $n=1$. Then $\Omega_\mu=\Gamma_\mu^L=1$ and~\eq{quotientU}
becomes trivial. The results of Ref.~\cite{GO83b} are reproduced
in this case as $N\ra\infty$ at finite $a$
since $\theta\ra\infty$ according to \eq{theta}.  
This limit is associated with the 't~Hooft limit of 
large-$N$ Yang--Mills theory
where only planar diagrams survive.

Alternatively, one can approach the continuum limit of the usual TEK
(with \mbox{$n=1$)} keeping $\theta$ fixed which requires
$a\sim 1/\sqrt{m}=N^{-1/D}$ as $N\ra\infty$. The period
$l=am \sim \sqrt{m}=N^{1/D}\ra\infty$ in this limit so that
a noncommutative gauge theory on $\IR^D$ is reproduced~\cite{AIIKKT}.

For $n>1$ (that is TEK with the quotient condition),
the noncommutativity parameter
\be 
\theta _{\mu\nu} = \frac{l^2}\pi \frac{n}{m} \varepsilon  _{\mu\nu}
\ee
can be kept finite as $N\ra\infty$ even for a finite $l$
if the dimensionless noncommutativity parameter
 $\Theta=n/m$ is kept finite. This means that the noncommutative
theory lives on a torus~\cite{CDS98}. The case of finite $N$ 
corresponds~\cite{ANMS99}
to the noncommutative lattice gauge theory~\rf{NCLGT}
which is a lattice regularization of the continuum theory.
Since the spatial extent of the lattice $l=am$ is finite,
one gets the relation
\be
p_{\rm max} \theta p_{\rm min}= 2\pi n
\label{UVIR}
\ee
between the UV cutoff $p_{\rm max}=\pi/a$ and the IR cutoff
$p_{\rm min}=2\pi/l$ which is similar to that discovered
in Ref.~\cite{uvir} for $\IR^4$ (associated with $n=1$ as
is discussed in the previous paragraph).

\newsection{Morita equivalence}

The continuum noncommutative gauge theory with rational values of
the dimensionless noncommutativity parameter
 $\Theta$ has an interesting property known as Morita equivalence~\cite{AS}.
We shall describe it for the lattice regularization 
associated with the simplest twist~\rf{twist} assuming
that the ratio $m/n=\tilde p$ is an integer.
Then the  noncommutative $U_{\theta}(1)$ gauge theory on
a $m^D$ periodic lattice is equivalent to ordinary $U(p)$
Yang--Mills theory with $p=\tilde p^{D/2}$ on a $n^D$ lattice
with twisted boundary conditions (representing the 't~Hooft flux)
and the coupling constant $g^2=e^2/p$ (where $e^2$ is the coupling
of the $U_{\theta}(1)$ theory).

The twisted boundary conditions read
\be
\tilde V_\mu(\tilde x+an\hat\nu)=
\tilde \Gamma_\nu \tilde V_\mu(\tilde x)\tilde \Gamma_\nu^\dagger
\label{tbc}
\ee
where $\tilde \Gamma_\nu$ are $p\times p$ twist eaters obeying
\be
\tilde \Gamma_\mu\tilde \Gamma_\nu= \tilde Z_{\mu\nu}
\tilde \Gamma_\nu \tilde \Gamma_\mu\,,~~~~
\tilde Z_{\mu\nu}= \e^{4\pi i \varepsilon_{\mu\nu}/\tilde p}
\ee 
($\tilde p$ is also assumed to be odd).
Here $\tilde Z_{\mu\nu}\in \IZ_p$ is not removable since $\tilde \Gamma_\mu$
are $SU(p)$ matrices. It represents the non-Abelian 't~Hooft flux.

We have discussed in the previous Section the equivalence of TEK (with 
the quotient condition in general) with $N=(mn)^{D/2}$
and the noncommutative $U_\theta(1)$ gauge theory on $\IT^D_m$.
Both theories have same $m^D$ degrees of freedom which are described
either by \eq{tildeU} or \eq{U(x)}. In the matrix language,
the noncommutativity emerges since
\be
J^n_k J^n_q= J^n_{k+q} \e^{2\pi i \frac nm k_\mu \varepsilon_{\mu\nu}q_\nu}
\label{JJn}
\ee
as it follows from the general \eq{JJ} for the given simplest twist.
In the noncommutative language, the noncommutativity resides in 
the star-product
\be
\e^{2\pi i \frac{kx}{l}}\star\e^{2\pi i \frac{qx}{l}}=
\e^{2\pi i \frac{(k+q)x}{l}}
\e^{2\pi i \frac nm k_\mu \varepsilon_{\mu\nu}q_\nu}
\label{een}
\ee
as it follows from the definition~\rf{lstar}.

When $m=\tilde p n$, a third equivalent model exists where the same dynamical
degrees of freedom are described by a $p\times p$ matrix field
\be
\tilde V^{ab}_\mu(\tilde x) 
= \sum_{k\in \IZs_m} \tilde J_k^{ab} 
\e^{2\pi i \frac{k\tilde x}{\tilde p a n}} U_\mu(k) \,.
\label{tildeV}
\ee
Here 
\be
\tilde J_k = \prod _\mu \tilde \Gamma_\mu ^{k_\mu} 
\e^{2\pi i \frac1{\tilde p}\sum_{\mu>\nu} \varepsilon_{\mu\nu}k_\mu k_\nu} 
\label{tildeJ}
\ee
similar to \eq{defJ}. The number of degrees of freedom $n^D p^2= m^D$
for $p=\tilde p^{D/2}$. The noncommutativity now resides in the matrix
part rather than the $\tilde x$-dependent part since
\be
\tilde J_k \tilde J_q= \tilde J_{k+q} 
\e^{2\pi i \frac 1{\tilde p} k_\mu \varepsilon_{\mu\nu}q_\nu}\,.
\label{tildeJJn}
\ee

The action of the third model is just the ordinary Wilson's lattice action
\be
S= \frac{p}{2e^2} \sum_{\tilde x\in \tilde \ITs{}^D_n} \sum_{\mu\neq\nu}
\tr_{(p)}{} \tilde V_\mu(\tilde x)
\tilde V_\nu(\tilde x+a\hat \mu)
\tilde V^\dagger_\mu(\tilde x+a\hat \nu)\tilde V^\dagger_\nu(\tilde x)\,.
\ee
The field $\tilde V_\mu(\tilde x)$ is quasi-periodic on $\tilde \IT{}^D_n$
and obeys the twisted boundary conditions~\rf{tbc} since
\be
\tilde \Gamma_\mu \tilde J_k \tilde \Gamma^\dagger_\mu= \tilde J_k 
\e^{2\pi i k_\mu /\tilde p}
\ee

For $n=1$ when $\tilde p=m$ and $p=N$, the third model 
lives on a unit hypercube
with twisted boundary conditions and coincides with
TEK. It is seen after the change of variables 
$U_\mu = \tilde V_\mu \tilde \Gamma_\mu$. 
In fact this formulation was the original motivation of Ref.~\cite{GO83b}
for TEK. Therefore, the derivation of noncommutative gauge theories
from TEK is a simplest example of Morita equivalence. 

In the continuum limit
($N\ra\infty$) when TEK is formulated via operators, the noncommutative
$U_\theta(1)$ gauge theory lives on $\IT^D$ 
with period $l$
and is Morita equivalent
at a rational value of $\Theta$ to the ordinary $U(p)$ gauge theory
on a smaller torus $\tilde \IT{}^D$ with twisted boundary conditions and 
$\tilde l =l/\tilde p$. The lattice regularization makes these results
rigorous~\cite{ANMS00a}. Arbitrary  (irrational or rational) value of $\Theta$ 
 can be obtained for the most general twist.
As is also shown in Ref.~\cite{ANMS00b},
the theories on tori can be obtained from TEK by choosing a
sophisticated twist instead of imposing the quotient condition.

\newsection{Fundamental matter \protect{\cite{ANMS00a}}}

The results of two previous Sections can be extended to the presence
of matter. 
Let $\phi(x)$ 
is a scalar matter field in the fundamental
representation of $U_\theta(1)$. The matter part of the action is
\be
S_{\rm matter}= -\sum_{x,\mu} 
\phi^* (x) \star \U_\mu(x) \star \phi(x+a\hat\mu)
+ M^2 \sum_x  \phi^* (x) \phi(x)
\label{Smatter}
\ee
and is invariant under the star-gauge transformation
\be
\phi(x)\ra \omega(x) \star\phi(x)\,,~~~~~~~
\phi^*(x)\ra \phi^*(x) \star\omega^*(x)
\ee
and \eq{stargauge} for $\U_\mu(x)$. 

At a rational value of $\Theta$, the action~\rf{Smatter} on a torus
is Morita equivalent to 
\be
S_{\rm matter}= -\sum_{\tilde x,\mu} 
\tr_{(p)} {}\Phi^* (\tilde x) \tilde V_\mu(\tilde x) 
\Phi(\tilde x+a\hat\mu)
+ M^2 \sum_{\tilde x} \tr_{(p)} {} \Phi^* (\tilde x) \Phi(\tilde x)
\label{Smattermatrix}
\ee
where the notations of the previous Section are used and
the $p\times p$ matrix field $\Phi^{ij}(\tilde x)$ 
obeys the twisted boundary conditions
\be
\Phi(\tilde x + \tilde l \hat \nu)= \tilde \Gamma_\nu \Phi(\tilde x )
\tilde \Gamma^\dagger_\nu
\ee
similar to~\eq{tbc} for the gauge field. The index $i$ of $\Phi^{ij}$
plays the role of color while $j$ plays the role of flavor (labeling
species). The color symmetry is local while the flavor symmetry
is global.
In particular, the model~\rf{Smattermatrix} reduces for $n=1$ to 
TEK for fundamental matter of Ref.~\cite{Das83}.

The continuum limit of the above formulas is obvious. 
The continuum $U_\theta(1)$ gauge theory with
fundamental matter (noncommutative QED) is reproduced as $N\ra\infty$.
For $\theta\ra\infty$ it is equivalent to large-$N$ QCD on $\IR^D$
in the
the Veneziano limit when the number of flavors of fundamental matter
is proportional to the number of colors 
so the matter survives in the large-$N$ limit. Again these results 
are rigorous since they are obtained in a regularized theory.

\newsection{Wilson loops in NCYM}

Observables in ordinary Yang--Mills theory can be expressed via
the Wilson loops. The standard way to derive proper formulas is to 
integrate over matter fields doing the Gaussian path integral. 
This strategy can be repeated for noncommutative gauge theory
with fundamental matter
given by the action~\rf{Smatter}. On the lattice one can use the 
large-mass expansion in $1/M^2$. We describe in this Section
what kinds of Wilson loops then emerge.

A lattice contour $C$ consisting of $J$ links
is defined by the set of unit vectors $\hat \mu_j$
associated with the direction of each link $j$ ($j=1,\ldots,J$)
forming the contour.
The parallel transporter from the point $x$ to the point
$x+\ell$ ($\ell= a \sum_j \hat \mu_j$) along $C$ is given by
\be
\U_\mu(x;C)=\U_{\mu_1}(x)\star\U_{\mu_2}(x+a\hat \mu_1) \star
\cdots \star \U_{\mu_J}(x+a \sum_{j=1}^{J-1} \hat \mu_j).
\ee 
It is star-gauge covariant:
\be
\U_\mu(x;C) \ra \omega(x)\star\U_\mu(x;C)\star\omega^*(x+\ell)
\label{starcovariant}
\ee
under the star-gauge transformation~\rf{stargauge}.

Given a function $S_\ell(x)$ with the property
\be
S_\ell(x)\star \omega(x) \star S^*_\ell(x) =\omega(x+\ell)
\label{SoS}
\ee
for arbitrary $\omega(x)$, it is easy to show that
\be
W(C)= \sum_x S_\ell(x)\star\U(x;C)
\label{defW}
\ee
is star-gauge invariant. The solution to \eq{SoS}  is
\be
S_\ell(x)=\e^{i \ell_\mu \theta^{-1}_{\mu\nu}x_\nu}\,,
\ee
where $\ell_\mu = an j_\mu$ with integer-valued $j_\mu$
(modulo possible windings). 

The continuum limit of \eq{defW} defines star-gauge invariant
Wilson loops in noncommutative gauge theory. In addition to
closed loops, there exist on $\IR^D$ open loops given by \eq{defW}
with an arbitrary value
of $\ell$~\cite{IIKK}. On $\IT^D$ 
the open Wilson loops are star-gauge invariant only for
 discrete values of $\ell$ measured in the units of 
$\pi \theta /l$~\cite{ANMS99}.
The closed Wilson loops appear in the sum-over-path representation
of the matter correlator $\LA \phi^*(x)\star\phi(x) \RA_\phi$
while the open Wilson loops appear for 
$\LA \phi^*(x)\star S_\ell(x)\star\phi(x+\ell) \RA_\phi$~\cite{ANMS00a}.
For integral $m/n=\tilde p$, the open Wilson loops in noncommutative
$U_\theta(1) $ gauge theory become the Polyakov loops winding around
$\tilde \IT{}^D$ in the Morita equivalent $U(p)$ Yang--Mills theory
with twisted boundary conditions.  

\newsection{D-brane interpretation~\protect{\cite{GM}}}

The results about the Morita equivalence have a simple interpretation
via T-duality in the D-brane language. Let us consider the case of 
a more general twist in $D=4$ when $SU(p)$ is decomposed as
$SU(p)\supset 1_{\tilde p_0} \otimes SU(\tilde p_1)  \otimes SU(\tilde p_2)$
($p= \tilde p_0\tilde p_1\tilde p_2$). The simplest twist~\rf{twist}  
above corresponds to $\tilde p_0=1$, $\tilde p_1=\tilde p_2=\tilde p$.

Let us consider the system of $\tilde p_0$ D3-branes populated by
$\tilde p_0\tilde p_1$ D1-branes localized in the 1,2-plane,
$\tilde p_0\tilde p_2$ D1-branes localized in the 3,4-plane,
and $p= \tilde p_0\tilde p_1\tilde p_2$ D-instantons. 
It is associated with noncommutative 
$U(\tilde p_0)$ gauge theory whose dimensionless noncommutativity
parameters equal~\cite{CO00}
\be
\Theta_{12}=\frac{\#D3}{\#D1}= \frac{1}{\tilde p_1}\,,~~~~~
\Theta_{34}=\frac{\#D3}{\#D1}= \frac{1}{\tilde p_2}\,.
\ee

After the T-duality transformation both in the 1,2 and 3,4 planes,
we get the system of $p$ D3-branes populated by
$\tilde p_0\tilde p_1$ D1-branes oriented in the 1,2-plane,
$\tilde p_0\tilde p_2$ D1-branes oriented in the 3,4-plane,
and $\tilde p_0$ D-instantons. Now
\be
\tilde\Theta_{12}=\frac{\#D3}{\#D1}= {\tilde p_1}\,,~~~~~
\tilde\Theta_{34}=\frac{\#D3}{\#D1}= {\tilde p_2}
\ee
represent magnetic fluxes in the ordinary $U(p)$ gauge theory.
The period matrix becomes
$\tilde \Sigma = {\rm diag}\,\left(l/\tilde p_1,  l/\tilde p_2\right)$  
due to the fluxes.

It is interesting to note the presence of $\tilde p_0$ D-instantons 
in the T-dual theory. They provide vanishing of the topological
charge $Q=\#D3 \#D(-1) - \#D1 \#D1=0$ as it should for the given twist
to provide zero-action vacuum configuration.

It is not clear how to extend the interpretation of this Section 
to the case when the fundamental matter is incorporated. 


\subsection*{Acknowledgments}

This work was supported in part by the 
CRDF Award RP1--2108.

\eop

\end{document}